\documentclass[12pt]{article}

\usepackage{moreverb,url}
\usepackage{enumitem, bm, amssymb, amsmath, multirow,setspace,titlesec}
\usepackage[table,xcdraw]{xcolor}
\usepackage{epsfig, amsfonts, amsbsy}
\usepackage{mathrsfs}
\usepackage{pdflscape}

\newcommand{\beginsupplement}{%
 \setcounter{table}{0}
 \renewcommand{\thetable}{S\arabic{table}}%
 \setcounter{figure}{0}
 \renewcommand{\thefigure}{S\arabic{figure}}%
 }

\usepackage{geometry}
\usepackage[super]{cite}

\usepackage[document]{ragged2e}

\usepackage{setspace}

\usepackage[colorlinks,bookmarksopen,bookmarksnumbered,citecolor=blue,urlcolor=red]{hyperref}

\usepackage{pdflscape}

\begin{document}

\title{Combining Real-World and Randomized Control Trial Data Using Data-Adaptive Weighting via the On-Trial Score}

\author{Joanna Harton$^1$, Brian Segal$^2$, Ronac Mamtani$^3$, \\ Nandita Mitra$^{1*}$, and Rebecca A. Hubbard$^{1*}$}

\date{}

\maketitle

\footnotetext[1]{Department of Biostatistics, Epidemiology, and Informatics, University of Pennsylvania, Philadelphia, PA, USA}
\footnotetext[2]{Indigo Ag, Boston, MA, USA}
\footnotetext[2]{Penn Medicine, Philadelphia, PA, USA}
\renewcommand{\thefootnote}{\fnsymbol{footnote}}
\footnotetext[1]{Co-senior authors}

\footnotetext{\textbf{Corresponding author:} Joanna Harton, Department of Biostatistics, Epidemiology, and Informatics, 423 Guardian Drive, Philadelphia, PA 19104, USA. \textbf{Email:} jograce@pennmedicine.upenn.edu}


\section*{Abstract}
Clinical trials with a hybrid control arm (a control arm constructed from a combination of randomized patients and real-world data on patients receiving usual care in standard clinical practice) have the potential to decrease the cost of randomized trials while increasing the proportion of trial patients given access to novel therapeutics. However, due to stringent trial inclusion criteria and differences in care and data quality between trials and community practice, trial patients may have systematically different outcomes compared to their real-world counterparts. We propose a new method for analyses of trials with a hybrid control arm that efficiently controls bias and type I error. Under our proposed approach, selected real-world patients are weighted by a function of the ``on-trial score,'' which reflects their similarity to trial patients. In contrast to previously developed hybrid control designs that assign the same weight to all real-world patients, our approach upweights of real-world patients who more closely resemble randomized control patients while dissimilar patients are discounted. Estimates of the treatment effect are obtained via Cox proportional hazards models. We compare our approach to existing approaches via simulations and apply these methods to a study using electronic health record data. Our proposed method is able to control type I error, minimize bias, and decrease variance when compared to using only trial data in nearly all scenarios examined. Therefore, our new approach can be used when conducting clinical trials by augmenting the standard-of-care arm with weighted patients from the EHR to increase power without inducing bias.

\section{Introduction}
Randomized clinical trials are the gold-standard for testing a new treatment though there can be some potential disadvantages to running a traditional clinical trial. Clinical trials for rare diseases can take a long time to accrue patients due to the rarity of the disease, which makes the clinical trial and drug approval process take longer. Additionally, if a prior Phase II trial has shown superiority of the new treatment over the standard treatment, it may not be ethical to randomize patients in a 1:1 ratio, rather it may be preferred to use a 2:1 or 3:1 ratio favoring the intervention over the standard-of-care. This can result in lower power and the inability to detect an effect even if one truly exists.

It is therefore appealing to consider combining clinical trial data on patients receiving a novel treatment with data on patients receiving the control therapy derived from electronic health records (EHR). \cite{ventz_design_2019} The appeal of including external patients receiving the standard-of-care is that it reduces or even completely eliminates the need to randomize patients to a control arm in the current trial. \cite{schmidli_beyond_2020} In a trial with an external control arm, all data on the standard-of-care is derived from EHR, while, in a hybrid control arm external patients receiving the standard-of-care from an EHR are combined with randomized trial patients in the control arm. \cite{burcu_real-world_2020} 

Electronic health records (EHR) contain a vast amount of data that can be relatively easily leveraged for research. These data are by nature observational, and, as such, there are a few key features of EHR data that are worth noting. First, EHR were developed for clinical care and billing purposes. Therefore, some of the information that researchers may be interested in may not be collected or may be contained in narrative text notes, which are difficult to analyze. Furthermore, the healthcare system in the United States, as well as in many other countries, is fragmented such that a patient's medical records may be spread across the databases of multiple healthcare systems, which can result in an incomplete or inadequate picture of a patient's health when relying on data from an individual EHR database.

Though conducting research with EHR data has distinct challenges, approaches have been developed to make beneficial use of this data source.

In 1976, Pocock proposed a method to combine randomized patients receiving the standard-of-care from historical clinical trials with intervention arm patients from a new trial to address the fact that many studies of the day examining the efficacy of a new treatment did not contain a randomized control arm, which made it difficult to draw causal conclusions. \cite{pocock_combination_1976} The approach represents a hybrid control arm combining patients randomized to the trial control arm with patients from the control arm of a historical trial. Pocock proposed six criteria for evaluating what constitutes an acceptable historical control arm as well as how much weight to assign to historical control patients relative to randomized control patients. \cite{pocock_combination_1976}

Pocock's six criteria can also be applied to the use of EHR data to construct the hybrid control arm of a trial with varying success. These six proposed guidelines along with considerations for application to an EHR-derived hybrid control arm are presented below: 
\begin{enumerate}
\item \emph{``Such a group must have received a precisely defined standard treatment which must be the same as the treatment for the randomized controls."} Patients derived from EHR data can be selected such that they are receiving the same primary treatment as the randomized trial control patients. However, supportive care and the care environment may differ between the trial and routine clinical practice.
\item \emph{``The group must have been part of a recent clinical study which contained the same requirements for patient eligibility."} By definition the EHR patients are not part of a recent clinical trial but efforts should be made to identify an EHR-derived cohort using eligibility criteria as similar to the trial as possible. Due to limitations of EHR data capture, it may be difficult to apply all clinical trial inclusion/exclusion criteria. \cite{ramsey_using_2020}
\item \emph{``The methods of treatment evaluation must be the same."} This criterion may or may not be met, depending on the outcome and method of outcome ascertainment used in the trial. While it is imperative for the outcome measure to be the same, outcome ascertainment may differ between the trial and EHR data capture. For example, even an outcome such as death can have sensitivity between $30\%$ and $90\%$, meaning that there are deaths that are not recorded in the EHR. \cite{carrigan_evaluation_2019}
\item \emph{``The distributions of important patient characteristics in the group should be comparable with those in the new trial."} Patients receiving the control treatment in routine care may differ in many ways from patients participating in a trial. The requirement of the same distribution of patient characteristics in the EHR patient pool as the trial patients is possible but highly unlikely to be exact; similarity, however, should be striven for.
\item \emph{``The previous study must have been performed in the same organization with largely the same clinical investigators."} This criterion is not able to be met by definition.
\item \emph{``There must be no other indications leading one to expect differing results between the randomized and historical controls."} This is unlikely to be met \textbf{completely} due to many differences between real-world and clinical trial care. However, if the EHR data is contemporaneous with the current trial it is reasonable to assume that care received by the EHR patients is similar to the care received by the trial control patients.
\end{enumerate}
Overall, if an EHR cohort can be constructed such that the patients are contemporaneous with the clinical trial, the same inclusion/exclusion criteria are applied to ensure comparable patient populations are under study, EHR control patients receive the same treatment as the clinical trial control arm patients, and the outcome and method of outcome ascertainment are as similar between EHR data capture and the trial as possible, it may be appropriate to use EHR data as part of a hybrid control arm.

There are several approaches to using external control patients when estimating treatment effects, including Pocock's approach which was among the first. Although these methods were developed in the context of historical controls, where control patients are drawn from a previously conducted clinical trial, they can also be applied to the case where control patients are drawn from an EHR database. This does not affect the implementation of the methods, though interpretation must be made carefully. We assume throughout this paper that only patients in the trial receive the intervention therapy. Pocock's method assumes that the parameter of interest is the difference between the mean outcome in the intervention arm patients and the mean outcome in the standard-of-care arm patients in the clinical trial. Pocock's method also assumes that the true mean value for the trial standard-of-care patients follows a normal distribution centered at a weighted sum of the sample mean of the trial standard-of-care patients and the sample mean of the external standard-of-care patients, where weights are selected based on the extent to which external standard-of-care patients are believed to be representative of trial standard-of-care patients and with a standard deviation also dependent upon these factors. \cite{pocock_combination_1976}

Chen, et al. introduced a Bayesian approach to estimating the effect of a treatment when using hybrid control arms that relies on power priors. \cite{chen_power_2000, ibrahim_power_2000} This approach incorporates external standard-of-care arm data with the current trial data by taking only a fraction of the information from each external standard-of-care patient. \cite{chen_power_2000, ibrahim_power_2000} The power prior can estimate many different estimands, including differences in means or proportions, hazard ratios, or odds ratios. In this method, the pool of external standard-of-care patients is weighted as a whole and the external standard-of-care patients are assigned anywhere from $0\%$ to $100\%$ of the weight that a current trial participant, whether intervention or standard-of-care, receives in the final model. \cite{chen_power_2000, ibrahim_power_2000} When $\alpha$, the weight assigned to external standard-of-care patients, is $0$, the power prior approach is equivalent to using no data from the external standard-of-care patients, and when $\alpha$ is $1$, the power prior approach is the same as fully pooling the external standard-of-care patients with the current trial data. \cite{chen_power_2000, ibrahim_power_2000} This method may be more interpretable than Pocock's method as the amount of information incorporated is quantified directly through $\alpha$ rather than through a variance parameter. \cite{chen_power_2000, ibrahim_power_2000} Similar to Pocock's method, the amount of information incorporated from the external standard-of-care patients must be prespecified by the researcher and sensitivity analyses are recommended to determine the robustness of results to choice of $\alpha$. \cite{chen_power_2000, ibrahim_power_2000, pocock_combination_1976}

Duan and Ye in 2008 \cite{duan_normalized_2008} and Neuenschwander, et al. in 2009 \cite{neuenschwander_note_2009} concurrently developed the normalized power prior approach, which extended the power prior model to estimate $\alpha$ from the data rather than using a prespecified $\alpha$. \cite{duan_normalized_2008, neuenschwander_note_2009} The normalized power prior approach allows for more weight to be allocated to the external standard-of-care patients when the external standard-of-care patients are similar in terms of outcome to the current trial patients and less weight when they are dissimilar. \cite{duan_normalized_2008, neuenschwander_note_2009}

These earlier methods focused on weighting external standard-of-care patients as a group, whether the weight is pre-specified or data-driven. More recent extensions have considered individual-weighting of external standard-of-care patients based on the similarity of each individual to patients in the trial standard-of-care arm. One such adaptation of the power prior approach proposes dividing the external standard-of-care patients into subgroups based on their similarity to the trial patients and assigning a weight to each subgroup. \cite{wang_propensity_2019} Another recently proposed method uses a modification of the propensity score, called the on-trial score, to create matches between the external standard-of-care patients and the trial standard-of-care patients in order to create a hybrid standard-of-care arm consisting of patients most similar to those in the clinical trial intervention arm. \cite{lin_propensity-score-based_2019} 

In this paper, we propose a new data-adaptive weighting method that addresses the limitation of assigning a single weight to the entire group of external standard-of-care patients by assigning weights to each individual in the external standard-of-care arm based on similarity to trial patients using the on-trial score. The use of individualized weights helps to account for the fact that patients included in EHR databases may be more heterogeneous than patients included in clinical trials. The proposed approach incorporates more information from standard-of-care patients who are more similar to trial participants than those who are not. 

The structure of the paper is as follows: in section 2 we define notation, outline the existing approaches used for hybrid standard-of-care arms and introduce our proposed method for combining trial standard-of-care patients with external standard-of-care patients. In section 3, simulations are presented to assess the relative performance of our new method compared to existing methods. Section 4 applies all methods discussed to a clinical study for patients with metastatic castration-resistant prostate cancer comparing the standard-of-care treatment of prednisone with the new treatment of abiraterone acetate in conjunction with prednisone with with external standard-of-care patients from a pseudo EHR, and section 5 provides a summary and discussion.

\section{Methods}

\subsection{Notation}
We assume the existence of a trial of size $N_{Trial}$ and an external data source, e.g. an EHR database consisting of patients meeting comparable inclusion/exclusion criteria and receiving the same treatment as trial standard-of-care arm patients, of size $N_{External}$. Let $N = N_{Trial} + N_{External}$. Let each patient in each database have information on a set of $k$ covariates, $\mathbf{X}_{N x k} = \{\mathbf{X}_1, \mathbf{X}_2, \hdots, \mathbf{X}_k \}$. Additionally, let $\mathbf{R}$ be an indicator such that $\mathbf{R}_i$ takes the value of 1 if the $i^{th}$ patient is in the external data source and 0 if the $i^{th}$ patient is enrolled in the trial. Similarly, let $\mathbf{T}$ be a treatment indicator such that $\mathbf{T}_i$ takes a value of 0 if the $i^{th}$ patient receives the standard-of-care and 1 if the $i^{th}$ patient receives the intervention. In our numerical experiments below, we a have a time to event outcome variable, $\mathbf{Y} = min(\mathbf{F},\mathbf{C})$, where $\mathbf{F}$ is the time of the event of interest (failure time), $\mathbf{C}$ is the censoring time. Also, let $\mathbf{S}$ be a status indicator where $\mathbf{S}_i$ takes a value of 1 if $\mathbf{T}_i<\mathbf{C}_i$ and 0 otherwise. Extensions to outcomes of other variable types follow directly from the likelihood-based formulation below. We denote the data available for the set of external standard-of-care patients as $\mathbf{D_0} = \{\mathbf{X}, \mathbf{S}, \mathbf{Y}|\mathbf{R}_i=1 \}$, the set of randomized standard-of-care patients as $\mathbf{D_C} = \{\mathbf{X}, \mathbf{S}, \mathbf{Y}|\mathbf{R}_i=0, \mathbf{T}_i=0 \}$, the set of randomized intervention arm patients as $\mathbf{D_T} = \{\mathbf{X}, \mathbf{S}, \mathbf{Y}|\mathbf{R}_i=0, \mathbf{T}_i=1 \}$, and the set of all trial patients as $\mathbf{D} = (\mathbf{D_T}, \mathbf{D_C})$. We also let $\theta$ denote a target parameter of interest that represents treatment efficacy which could be parameterized as a difference in mean event times or hazard ratios comparing intervention arm and standard-of-care patients.

Below we summarize existing and proposed approaches to incorporating data from external standard-of-care patients into an analysis of treatment efficacy.

\subsection{Existing Approaches to Hybrid Control Trials}

\subsubsection{Na\"{i}ve Approaches}
Ignoring the external data and using only the current trial data serves as a positive control with regards to the minimum bias that can be attained when estimating the parameter of interest. In this method, only the data from the patients enrolled in the trial are analyzed and the patients from the external data source are left out.

Fully pooling the external data with the trial data serves as a negative control with regards to the amount of bias that is likely to occur when estimating the treatment effect. In this case, the external patients are given the same weight as the trial patients and all external patients are included in the analysis.

\subsubsection{Bayesian Approaches}
The power prior (PP) approach combines the external patients with the trial patients such that each external patient has a weight less than 1. \cite{ibrahim_power_2000} The power prior approach assigns the same weight, $\alpha$, such that $0<\alpha < 1$, to all patients in the external data source and the value of $\alpha$ is prespecified by the researcher. The power prior approach proposes the following prior distribution for $\theta$: $\pi(\theta | {\mathbf{D_0}}, \alpha) \propto \mathscr{L}(\theta | {\mathbf{D_0}})^\alpha \pi(\theta)$ which yields the following posterior distribution for $\theta$: $\pi(\theta | {\mathbf{D_0}}, \mathbf{D}, \alpha) \propto \mathscr{L}(\theta | \mathbf{D})[\mathscr{L}(\theta | {\mathbf{D_0}})^\alpha \pi(\theta)]$.

The normalized power prior approach (NPP) is similar to the power prior approach except that $\alpha$ is estimated from the data rather than being pre-specified by the researcher. \cite{duan_normalized_2008} The normalized power prior approach specifies a conditional prior for $\theta$ given $\alpha$ and a marginal distribution for $\alpha$. The normalized power prior approach has the form: 
\begin{equation*}
\pi(\theta, \alpha | {\mathbf{D_0}}) \propto \frac{\mathscr{L}(\theta | {\mathbf{D_0}})^\alpha \pi(\theta) }{\int_\Theta \mathscr{L}(\theta | {\mathbf{D_0}})^\alpha \pi(\theta) d\theta} \pi(\alpha),
\end{equation*}
which results in the following posterior: 
\begin{equation*}
\pi(\theta, \alpha | {\mathbf{D_0}}, \mathbf{D}) \propto \mathscr{L}(\theta | \mathbf{D}) \pi(\theta, \alpha | {\mathbf{D_0}}) \propto \frac{\mathscr{L}(\theta | \mathbf{D})\mathscr{L}(\theta | {\mathbf{D_0}})^\alpha \pi(\theta) \pi(\alpha)}{\int_\Theta \mathscr{L}(\theta | {\mathbf{D_0}})^\alpha \pi(\theta) d\theta}. 
\end{equation*}

If $\pi(\alpha)$ is proper then the normalized power prior will also be proper.

\subsubsection{Lin's Method}
Lin's method uses an on-trial score, similar to a propensity score, where the outcome of interest is inclusion in the trial to construct a matched set of external standard-of-care patients and weight their likelihood contribution. \cite{lin_propensity-score-based_2019} The on-trial score is estimated as the probability that a patient is in the clinical trial given their baseline covariates using a logistic regression model. Next, optimal pair matching is performed using the on-trial score so that each trial patient receiving the intervention is matched with an external standard-of-care patient. The selected external standard-of-care patients form a pool from which $N_T-N_C$ patients are randomly drawn so that the augmented trial has a 1:1 ratio of treated to standard-of-care patients. In the outcome model the external standard-of-care patients are weighted by their on-trial scores while the trial patients are given a weight of one. \cite{lin_propensity-score-based_2019}

\subsection{Proposed Approach: Data-Adaptive Weighting}

In our proposed approach, we let the on-trial score be defined as the probability that the $i^{th}$ patient is included in the trial given the observed baseline covariates, $P(R_i=0|\mathbf{X}_i) = e(\mathbf{X}_i)$. To maximize the similarity between external and randomized standard-of-care patients, we then limit the set of external standard-of-care patients to the subset with the highest on-trial scores, such that the number of external standard-of-care patients selected results in a hybrid control arm of the same size as the intervention arm. Let $D^*_0$ represent data for the subset of $D_0$ with the $N_{D_T}-N_{D_C}$ largest on-trial scores. The on-trial scores are then transformed to obtain values for $\hat{\gamma}_i$ such that $\hat{\gamma}_{i} = e(\mathbf{X}_i)/(1-e(\mathbf{X}_i))$ and standardized such that $\hat{\gamma}_i^* = \hat{\gamma}_i N_{\mathbf{D^*_0}}/\sum_{i=1}^{N_{\mathbf{D^*_0}}} \hat{\gamma}_i$. The inverse odds weight is used as we are interested in the average treatment effect on the treated, or in this case, the average treatment effect for those on-trial. This weighting method assigns all trial patients their full weight and only up- or down-weights the selected external standard-of-care patients.

Estimation for data-adaptive weighting then uses a prior for $\theta$ of the form: 
\begin{equation}
\pi(\theta | \bm{\hat{\gamma}^*}, {\mathbf{D^*_0}}) \propto \prod_{i=1}^{N_{\mathbf{D^*_0}}} \big[ \mathscr{L}(\theta | {{\mathbf{D^*_0}}_i})^{\hat{\gamma}^*_i} \big] \pi(\theta),
\end{equation}
which gives the following posterior: 
\begin{equation}
\pi(\theta | \bm{\hat{\gamma}^*}, {\mathbf{D^*_0}}, \mathbf{D}) \propto \prod_{j=1}^{N_{\mathbf{D}}} \big[ \mathscr{L}(\theta | \mathbf{D}_j) \big] \prod_{i=1}^{N_{\mathbf{D^*_0}}} \big[ \mathscr{L}(\theta | {{\mathbf{D^*_0}}_i})^{\hat{\gamma}^*_i} \big] \pi(\theta)
\label{daw_posterior}
\end{equation}

We note that all patients are used in the estimation of the on-trial score as, assuming trial patients are randomly assigned to intervention arm, the distribution of baseline covariates is the same for standard-of-care arm and intervention arm patients. Here the on-trial score is estimated via a logistic regression, rather than being jointly estimated with $\theta$. However, the on-trial score may be estimated using more flexible modeling such as a random forest or ensemble machine learning if desired.

\subsection{Estimation}

The Bayesian estimation approach for the models presented above, under certain conditions, can be approximated using a frequentist analog. For example, in the case of the DAW method, when a non-informative prior is used for $\theta$, the prior for DAW, $\pi(\bm{\theta} | \mathbf{D_0}, \bm{\hat{\gamma}^*}) \propto \big[ \mathscr{L}(\bm{\theta} | \mathbf{D_0})^{\hat{\gamma}_i^*} \big] \pi(\bm{\theta})$, is equivalent to $\pi(\bm{\theta} | \mathbf{D_0}, \bm{\hat{\gamma}^*}) \propto \mathscr{L}(\bm{\theta} | \mathbf{D_0})^{\hat{\gamma}_i^*}$. In this case the posterior mode corresponds to the maximum likelihood estimator for a weighted parametric survival model. In this case, estimates from a weighted Cox proportional hazards model with weights of 1 for trial patients and $\bm{\hat{\gamma}^*}$ for selected external standard-of-care patients will provide similar estimates to the Bayesian estimates with flat priors. This approach may be preferable to a fully Bayesian estimation approach because of its relative computational efficiency and insensitivity to prior specification. In numerical experiments below, we investigate performance of estimation using this weighted Cox approach for all methods described above.

Our proposed method, data-adaptive weighting (DAW), builds on the power prior approach and Lin's approach. While the power prior approach and normalized power prior approach both use the same $\alpha$ value for all patients, DAW uses individual weights for the external patients depending on their similarity to the trial patients. Lin's approach uses individual weights for each of the external patients. However, the selected external patients are based upon matching on the on-trial score and selected subjects are directly weighted by the on-trial score. 

\section{Simulation Study}
We conducted a simulation study to investigate the bias, efficiency, effective sample size, and type I error of the existing methods outlined above and the data-adaptive weighting method. Data were simulated with the objective of generating simulated data resembling a real-world study using trial data and external standard-of-care data from an EHR database. 

Data were simulated for four covariates ($\bm{X}$), a real-world indicator ($\bm{R}$), a treatment indicator ($\bm{T}$), a failure time ($\bm{T}$), and a censoring time ($\bm{C}$). We simulated data for trials of two different sizes ($100, 1000$), each with two different randomization ratios of the number of patients in the intervention arm to the number of patients in the standard-of-care arm (2:1 and 3:1). The number of external standard-of-care patients available was equal to the number of trial patients. These values were selected to mirror real-world scenarios for unbalanced clinical trials and provide enough potential EHR patients to distinguish between the performance of the various methods. The hazard ratio for failure for patients in the intervention arm versus randomized standard-of-care patients (treatment effect) was allowed to take values of $0.5, 0.75, 0.875,$ and $1$. Two different strengths of confounding of the relationship between baseline covariates and failure were also explored: mild confounding and strong confounding. We note that these covariates confound the relationship between enrollment in the trial and the outcome because the baseline covariate distribution differs between trial and external standard-of-care patients. Analyses limited to the trial population are unconfounded because there is no relationship between trial arm and baseline covariates due to randomization. Censoring rate was held constant across all simulation scenarios. External standard-of-care patients had censoring times arising from an exponential distribution with rate 0.4 and trial patients had censoring times arising from an exponential distribution with rate 0.1.

Relationships among the simulated variables and the complete set of distributions and parameter values used to simulate data, along with examples of EHR-derived covariates used to motivate the simulation study are provided in Table \ref{sims}.

\renewcommand{\arraystretch}{1.2}
\begin{table}[ht]
\caption{Data generation scheme and parameter values used for simulation study.}
\centering
\footnotesize
\begin{tabular}{|c|cc|c|}
\hline
\textbf{Variable} & \multicolumn{1}{c|}{\textbf{Trial Distribution}} & \textbf{External Distribution} & \textbf{Analogous Variable} \\ \hline
$X_1$ & \multicolumn{1}{c|}{$\text{Bernoulli}(0.5)$} & $\text{Bernoulli}(0.55)$ & Gender \\ \hline
$X_2$ & \multicolumn{1}{c|}{$\text{Bernoulli}(0.6)$} & $\text{Bernoulli}(0.4)$ & College Degree \\ \hline
$X_3$ & \multicolumn{1}{c|}{$\text{Normal}(60, 5)-60$} & $\text{Normal}(60, 10)-60$ & HDL Cholesterol \\ \hline
$X_4$ & \multicolumn{1}{c|}{$\text{Normal}(21, 2)-21$} & $\text{Normal}(23, 2)-21$ & BMI \\ \hline
T & \multicolumn{1}{c|}{$\text{Bernoulli}(t)$, $t=\{0.67, 0.75\}$} & 0 & Treatment Indicator \\ \hline
\multirow{2}{*}{$Y_{failure}$} & \multicolumn{2}{c|}{\begin{tabular}[c]{@{}c@{}}$\text{Exp}[log(\eta)T+log(\bm{\beta})\bm{X}]$\end{tabular}} & \multirow{2}{*}{Failure Time} \\
 & \multicolumn{1}{r}{\begin{tabular}[c]{@{}r@{}}$\eta =$\\  $\bm{\beta} =$\\ \textcolor{white}{.}  \end{tabular}} & \multicolumn{1}{l|}{\begin{tabular}[c]{@{}l@{}}$\{0.5, 0.75, 0.875, 1 \}$\\  $\{  (1.25, 0.67, 0.98, 1.06), $\\ $ (2.25, 0.4, 0.93, 1.21) \}$\end{tabular}} & \\ \hline
$Y_{censor}$ & \multicolumn{1}{c|}{$\text{Exp}(0.1)$} & $\text{Exp}(0.4)$ & Censoring Time \\ \hline
\end{tabular}
\label{sims}
\end{table}
\renewcommand{\arraystretch}{1}
 
In numerical examples, the on-trial score was estimated using logistic regression including all baseline covariates as predictors. Weighted Cox proportional hazards models with a treatment indicator as the sole covariate were fit as the outcome model to estimate the treatment effect. 

Each simulation scenario was repeated 1,000 times. We estimated bias relative to the true marginal treatment effect, empirical variance, 95\% confidence interval coverage probabilities, power (for scenarios with non-null treatment effects), and type I error (for scenarios with a null treatment effect). Power and type I error are estimated for hypothesis testing using a significance threshold of 0.05.

\begin{figure}[ht!]
    \centering
    \includegraphics[width=6in]{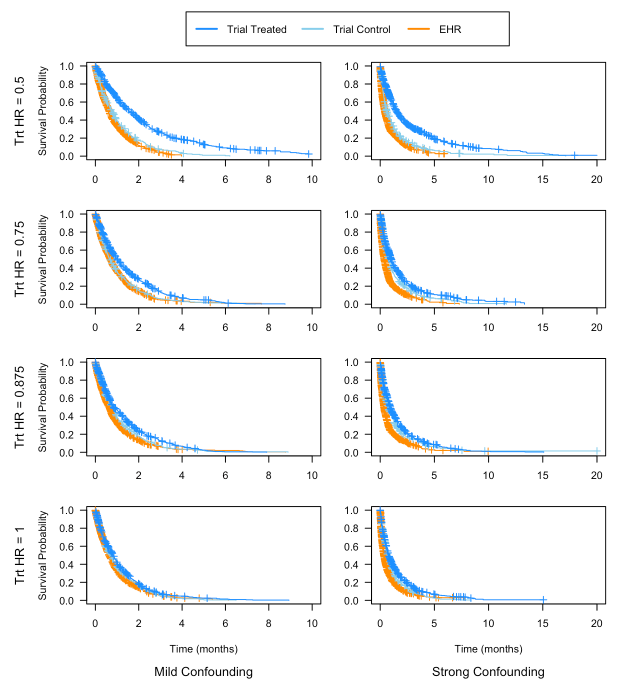}
    \caption{Kaplan-Meier curves for sample simulated datasets for each of four conditional treatment hazard ratios and two strengths of confounding.}
    \label{fig:sim_km}
\end{figure}

We first examined the performance of alternative methods as we varied the ratio of patients in the intervention arm to standard-of-care patients in the trial. In these analyses neither the proportion of trial patients in the intervention arm nor the number of EHR patients available as a function of the number of trial patients affected the pattern of the results. Therefore, only a 2:1 intervention arm to standard-of-care arm ratio with the same number of EHR patients available as trial patients are presented here. Sample Kaplan-Meier curves for each treatment hazard ratio and confounding level combination show the difference in survival over time across the three groups of patients (Supplemental Figure \ref{fig:sim_km}).

\subsection{Simulation Results}

\begin{figure}[ht!]
 \centering
 \includegraphics[width=6.5in]{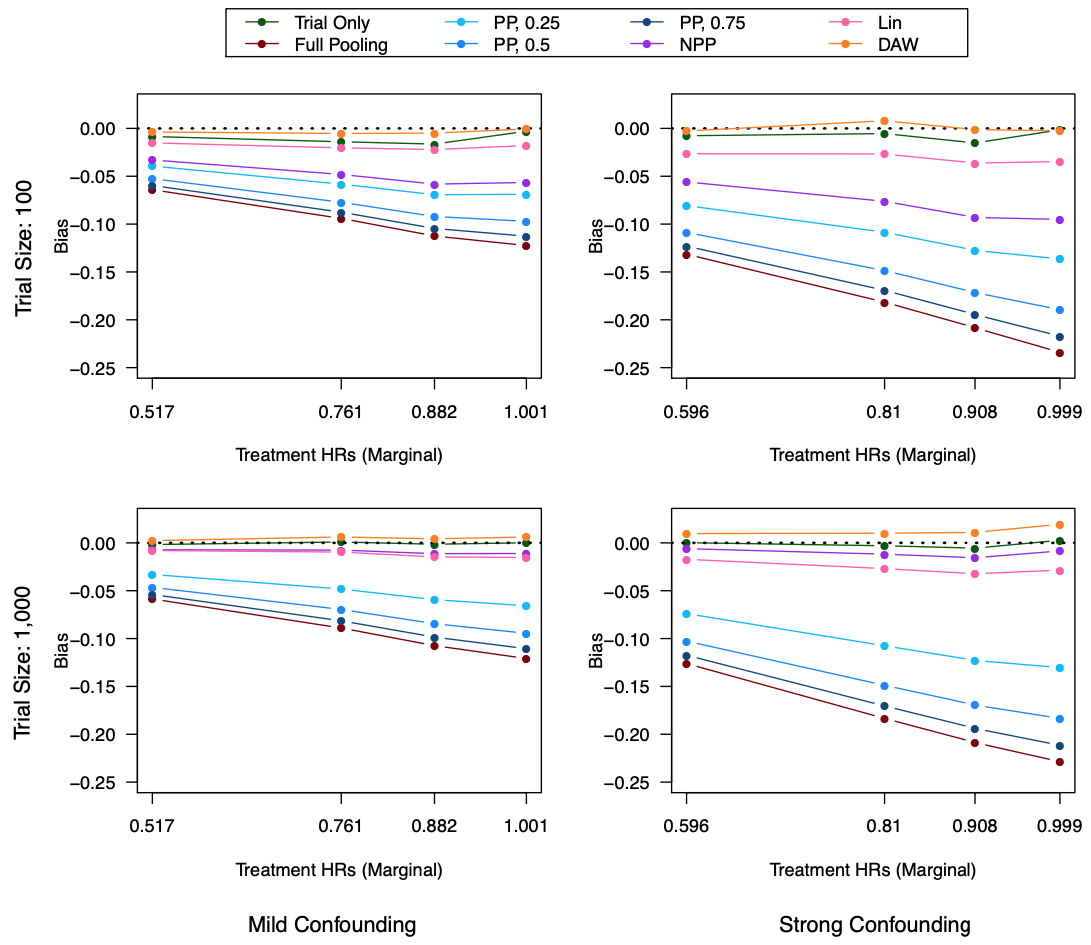}
 \caption{Bias for methods when applied to a trial with a 2:1 randomization ratio. PP = power prior, NPP = normalized power prior, DAW = data-adaptive weighting.}
 \label{fig:bias}
\end{figure}

\begin{figure}[ht!]
 \centering
 \includegraphics[width=6.5in]{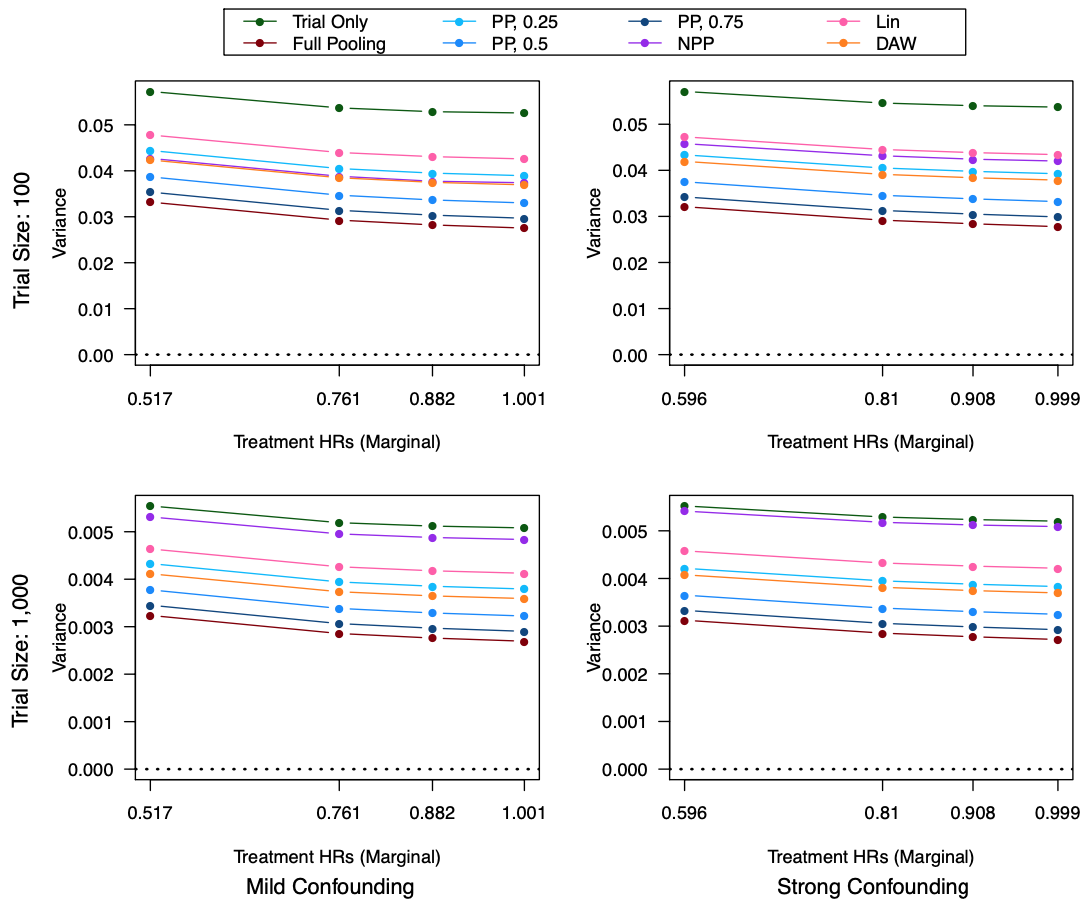}
 \caption{Variance for methods when applied to a trial with a 2:1 randomization ratio. PP = power prior, NPP = normalized power prior, DAW = data-adaptive weighting.}
 \label{fig:variance}
\end{figure}

\begin{figure}[ht!]
 \centering
 \includegraphics[width=6.5in]{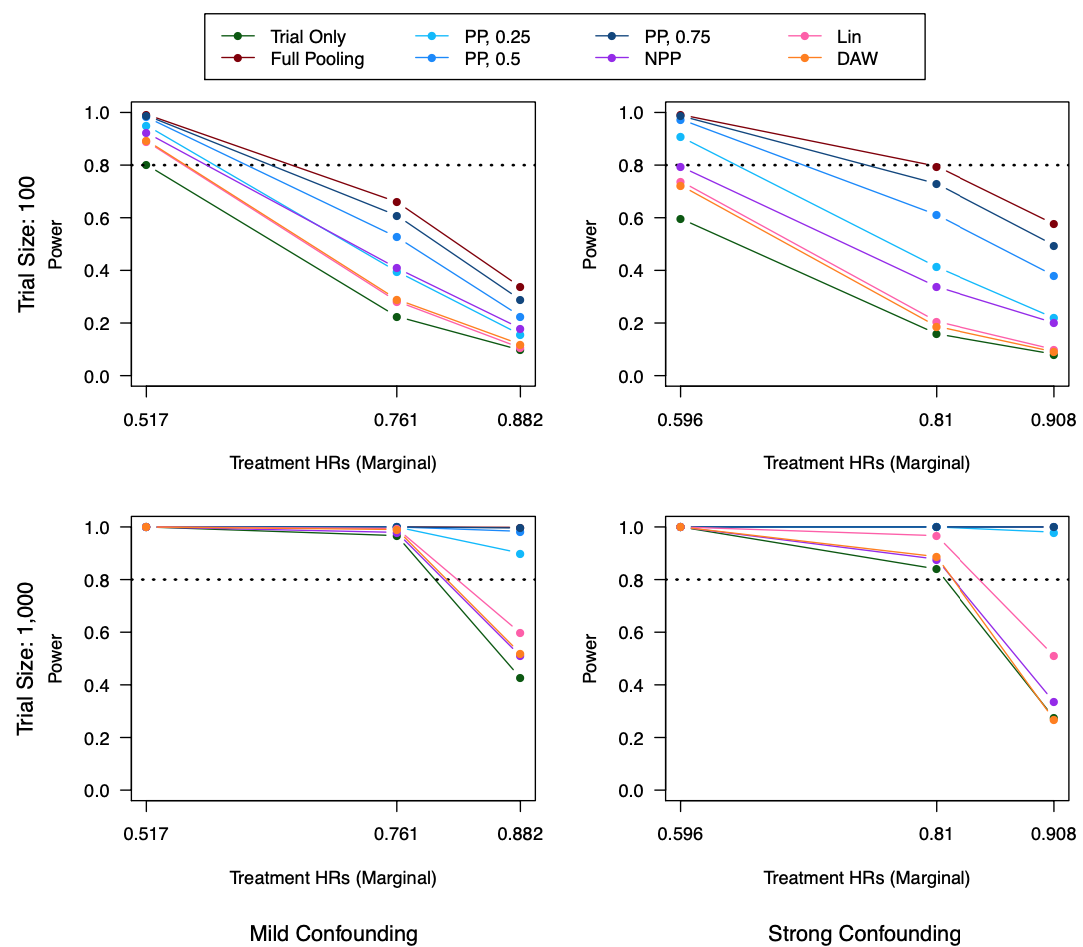}
 \caption{Power for methods when applied to a trial with a 2:1 randomization ratio. $80\%$ power is marked with a dotted line. PP = power prior, NPP = normalized power prior, DAW = data-adaptive weighting.}
 \label{fig:power}
\end{figure}

\begin{figure}[ht!]
 \centering
 \includegraphics[width=6.5in]{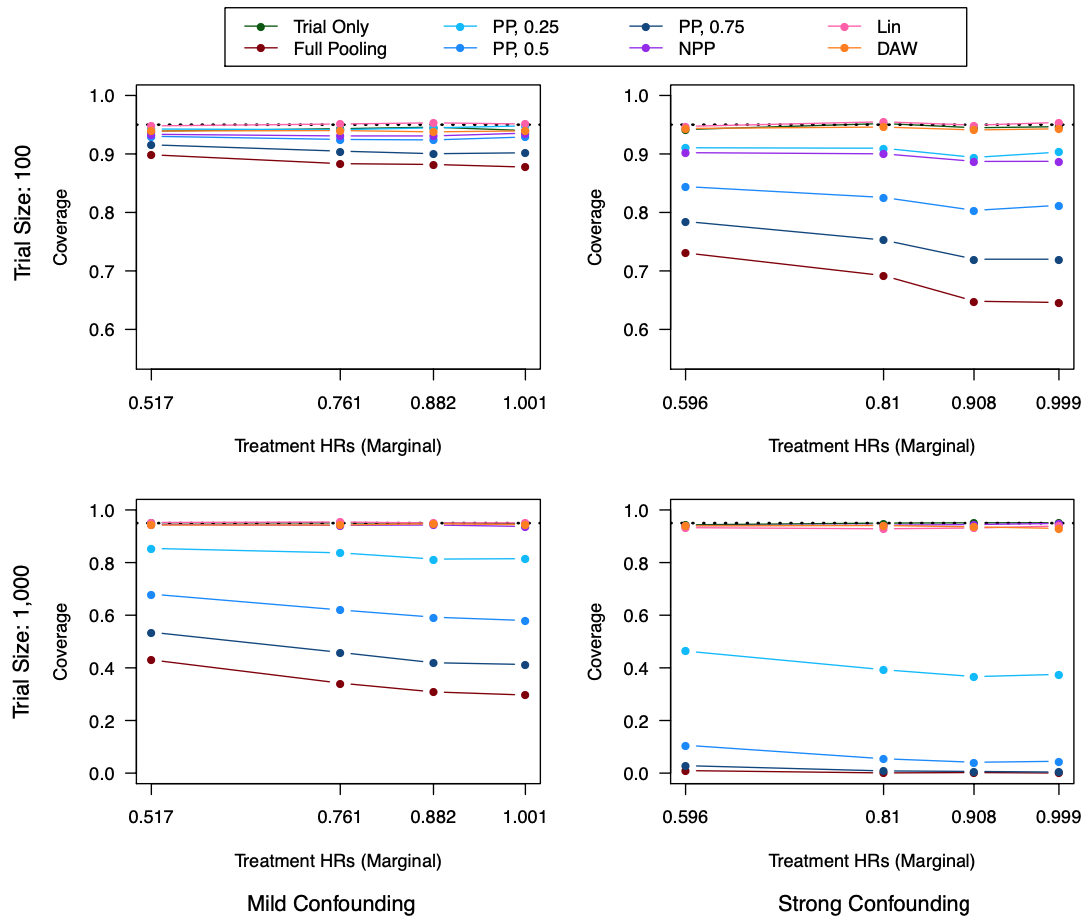}
 \caption{Coverage of the true treatment effect hazard ratio when a 95$\%$ confidence interval is used for methods when applied to a trial with a 2:1 randomization ratio. 95$\%$ coverage is marked with a dotted line. PP = power prior, NPP = normalized power prior, DAW = data-adaptive weighting.}
 \label{fig:coverage}
\end{figure}

\begin{figure}[ht!]
 \centering
 \includegraphics[width=6.5in]{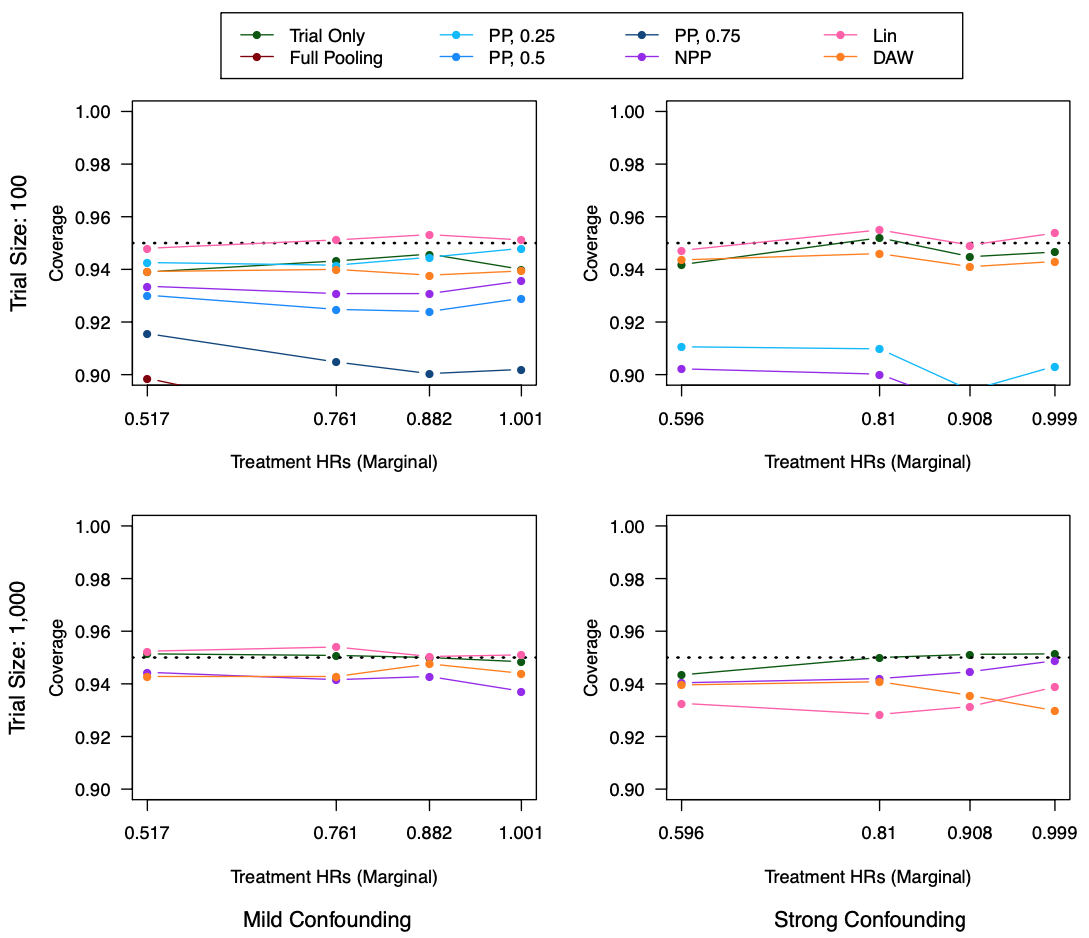}
 \caption{Zoomed-in view of coverage of the true treatment effect hazard ratio when a 95$\%$ confidence interval is used for methods when applied to a trial with a 2:1 randomization ratio. 95$\%$ coverage is marked with a dotted line. PP = power prior, NPP = normalized power prior, DAW = data-adaptive weighting.}
 \label{fig:coverageZOOM}
\end{figure}

The results of the simulation study show that fully pooling the data from the available EHR patients with the trial data results in large biases under all conditions examined (Figure \ref{fig:bias}). As expected, using only the trial patient data results in negligible bias. The power prior method was also substantially biased across all three $\alpha$ values explored, with $\alpha = 0.25$ having the least bias and $\alpha = 0.75$ having the most bias (Figure \ref{fig:bias}). The normalized power prior was biased when the trial size was 100 but displayed minimal bias when the trial size was 1,000. This is explained by the fact that $\hat{\alpha}$ was estimated to be approximately 0.22-0.36 for the trial size of 100 and 0.01-0.03 for the trial size of 1,000 (Table \ref{tab:alpha_npp}). Therefore, although the normalized power prior approach performed well in terms of bias this is because little information from the EHR patients was incorporated. Both Lin's method and DAW displayed low bias across all scenarios examined. DAW had consistently lower bias than Lin's method (Figure \ref{fig:bias}).

The variances of the estimates reflect the extent to which data from external standard-of-care subjects was incorporated. The trial only approach had the largest variance and full pooling of all EHR patients with the trial patients had the smallest variance (Figure \ref{fig:variance}). The power prior approach had variance between the trial only and full pooling methods, with variance inversely proportional to $\alpha$. The normalized power prior had substantially smaller variance than the trial only approach when the trial size was 100 due to the larger $\hat{\alpha}s$
(Figure \ref{fig:variance}, Table \ref{tab:alpha_npp}). DAW had smaller variance than Lin's method under all conditions examined. This is due to the larger effective sample size for any given scenario (Figure \ref{fig:variance}, Table \ref{tab:ess}).

DAW was able to achieve the targeted 1:1 intervention to standard-of-care ratio, while Lin's method resulted in substantially smaller effective sample sizes. DAW includes about twice as many external patients as Lin's method does in the scenarios examined (Table \ref{tab:ess}). 

The full pooling and power prior methods had high power due to their large effective sample sizes and biased treatment effect estimated which were biased away from the null; however, had the covariate effects been in the other direction the bias would have been towards the null and these methods would have had lower power (Figures \ref{fig:power}, \ref{fig:bias}). The normalized power prior approach had higher power relative to the trial only method when the trial size was 100 but not when the trial size is 1,000. Lin's method and DAW had higher power than the trial only approach and were quite similar to one another when the trial size was 100; Lin's method had slightly higher power than DAW when the trial size was 1,000 (Figure \ref{fig:power}).

Clearly, unless the strength of confounding and the trial size are small, full pooling of all EHR patients or using the power prior with one of the $\alpha$ values examined provides very poor coverage of the true HR (Figure \ref{fig:coverage}). NPP only provides nominal coverage when the trial size is 1,000. Both Lin's method and DAW had nominal coverage in all scenarios except under strong confounding when the trial size was 1,000; in that case Lin's method had approximately $93-94\%$ coverage and DAW had $94\%$ coverage except when the marginal treatment ratio is nearly 1, in which case the coverage dropped to $9\%$ (Figures \ref{fig:coverage}, \ref{fig:coverageZOOM}).

Type I error was controlled at the 5\% level when only the trial data were analyzed, and was poorly controlled under the full pooling and power prior methods (Table \ref{tab:type1error}). As expected, NPP controlled type I error when $\hat{\alpha}$ was small (i.e., sample size of 1,000) and when there was mild confounding (Tables \ref{tab:type1error}, \ref{tab:alpha_npp}). Both Lin's method and DAW controlled type I error under all scenarios examined except when there was strong confounding and a trial size of 1,000; type I error was slightly inflated in this case to around $6\%$ (Table \ref{tab:type1error}).

\renewcommand{\arraystretch}{1.2}
\begin{table}[ht]
\centering
\caption{Type I error for trials with a 2:1 randomization ratio. PP = power prior, NPP = normalized power prior, DAW = data-adaptive weighting.}
\label{tab:type1error}
\resizebox{0.9\textwidth}{!}{%
\begin{tabular}{|r|c|c|c|c|}
\hline
 & \multicolumn{2}{c|}{\textbf{Mild Confounding}} & \multicolumn{2}{c|}{\textbf{Strong Confounding}} \\ \hline
 & \textbf{Trial = 100} & \textbf{Trial = 1,000} & \textbf{Trial = 100} & \textbf{Trial = 1,000} \\ \hline
\textbf{Trial Only} & 0.05 & 0.051 & 0.052 & 0.046 \\ \hline
\textbf{Full Pooling} & 0.126 & 0.716 & 0.356 & 0.999 \\ \hline
\textbf{PP,} $\bm{\alpha = 0.25}$ & 0.053 & 0.177 & 0.098 & 0.619 \\ \hline
\textbf{PP,} $\bm{\alpha = 0.5}$ & 0.069 & 0.403 & 0.192 & 0.953 \\ \hline
\textbf{PP,} $\bm{\alpha = 0.75}$ & 0.097 & 0.583 & 0.287 & 0.996 \\ \hline
\textbf{NPP} & 0.061 & 0.061 & 0.117 & 0.047 \\ \hline
\textbf{Lin} & 0.049 & 0.046 & 0.044 & 0.060 \\ \hline
\textbf{DAW} & 0.052 & 0.048 & 0.050 & 0.059 \\ \hline
\end{tabular}%
}
\end{table}
\renewcommand{\arraystretch}{1}

\renewcommand{\arraystretch}{1.2}
\begin{table}[ht]
\centering
\caption{Effective sample size for trials with a 2:1 randomization ratio. PP = power prior, NPP = normalized power prior, DAW = data-adaptive weighting}
\label{tab:ess}
\resizebox{0.9\textwidth}{!}{%
\begin{tabular}{|r|c|c|c|c|}
\hline
 & \multicolumn{2}{c|}{\textbf{Mild Confounding}} & \multicolumn{2}{c|}{\textbf{Strong Confounding}} \\ \hline
 & \textbf{Trial = 100} & \textbf{Trial = 1,000} & \textbf{Trial = 100} & \textbf{Trial = 1,000} \\ \hline
\textbf{Trial Only} & 100 & 1000 & 100 & 1000 \\ \hline
\textbf{Full Pooling} & 200 & 2000 & 200 & 2000 \\ \hline
\textbf{PP,} $\bm{\alpha = 0.25}$ & 125 & 1250 & 125 & 1250 \\ \hline
\textbf{PP,} $\bm{\alpha = 0.5}$ & 150 & 1500 & 150 & 1500 \\ \hline
\textbf{PP,} $\bm{\alpha = 0.75}$ & 175 & 1750 & 175 & 1750 \\ \hline
\textbf{NPP} & 135 & 1033 & 123 & 1013 \\ \hline
\textbf{Lin} & 116 & 1166 & 116 & 1166 \\ \hline
\textbf{DAW} & 134 & 1340 & 134 & 1340 \\ \hline
\end{tabular}%
}
\end{table}
\renewcommand{\arraystretch}{1}

\renewcommand{\arraystretch}{1.2}
\begin{table}[ht]
\begin{center}
\caption{Mean $\alpha$ (alpha) values for normalized power prior approach for trials with a 2:1 randomization ratio. PP = power prior, NPP = normalized power prior, DAW = data-adaptive weighting}
\label{tab:alpha_npp}
\resizebox{0.9\textwidth}{!}{%
\begin{tabular}{|r|c|c|c|c|}
\hline
 & \multicolumn{2}{c|}{\textbf{Mild Confounding}} & \multicolumn{2}{c|}{\textbf{Strong Confounding}} \\ \hline
 & \textbf{Trial = 100} & \textbf{Trial = 1,000} & \textbf{Trial = 100} & \textbf{Trial = 1,000} \\ \hline
 \textbf{HR: 0.5} & 0.36 & 0.03 & 0.22 & 0.01 \\ \hline
 \textbf{HR: 0.75} & 0.36 & 0.03 & 0.23 & 0.01 \\ \hline
 \textbf{HR: 0.875} & 0.36 & 0.03 & 0.24 & 0.01 \\ \hline
 \textbf{HR: 1} & 0.35 & 0.03 & 0.23 & 0.01 \\ \hline
\end{tabular}%
}
\\
\end{center}
\footnotesize{Note: Hazard ratios listed are the conditional treatment hazard ratios as opposed to the marginal treatment hazard ratios.}
\end{table}
\renewcommand{\arraystretch}{1}

\section{Case Study: Metastatic Castration-Resistant Prostate Cancer} 

\renewcommand{\arraystretch}{1.2}
\begin{table}[ht!]
\centering
\caption{Characteristics of patients in the Janssen MCRPC cohort by data source}
\label{tab:mcrpc_tab1}
\resizebox{0.9\textwidth}{!}{%
\begin{tabular}{|r|c||c|c|}
\hline
\multicolumn{1}{|c|}{\textbf{}} & \multicolumn{1}{c||}{\textbf{}} & \multicolumn{2}{c|}{\textbf{Clinical Trial}} \\
 & \multicolumn{1}{c||}{\textbf{\begin{tabular}[c]{@{}c@{}}Pseudo\\ EHR\end{tabular}}} & \multicolumn{1}{c|}{\textbf{\begin{tabular}[c]{@{}c@{}}Standard-of-Care \\ Arm\end{tabular}}} & \multicolumn{1}{c|}{\textbf{\begin{tabular}[c]{@{}c@{}}Intervention \\ Arm\end{tabular}}} \\ 
\textbf{} & N = 1000 & N = 394 & N = 791 \\ \hline
\textbf{Age, median (IQR)} & 69 (63 - 76) & 69 (63 - 75) & 69 (64 - 75) \\ \hline
\textbf{ECOG PS, N (\%)} & & & \\
\textbf{0} & 244 (24.4) & 140 (35.5) & 262 (33.1) \\
\textbf{1} & 572 (57.2) & 209 (53.0) & 447 (56.5) \\
\textbf{2} & 184 (18.4) & 45 (11.4) & 82 (10.4) \\ \hline
\textbf{Gleason Score, N (\%)} & & & \\
\textbf{1} & 66 (6.6) & 0 (0.0) & 1 (0.1) \\
\textbf{2} & 11 (1.1) & 15 (3.8) & 31 (3.9) \\
\textbf{3} & 12 (1.2) & 1 (0.3) & 2 (0.3) \\ 
\textbf{4} & 4 (0.4) & 1 (0.3) & 3 (0.4) \\
\textbf{5} & 19 (1.9) & 3 (0.8) & 4 (0.5) \\
\textbf{6} & 78 (7.8) & 2 (0.5) & 24 (3.0) \\
\textbf{7} & 304 (30.4) & 32 (8.1) & 76 (9.6) \\
\textbf{8} & 152 (15.2) & 151 (38.3) & 286 (36.2) \\
\textbf{9} & 354 (35.4) & 76 (19.3) & 159 (20.1) \\
\textbf{10} & 0 (0.0) & 113 (28.7) & 205 (25.9) \\ \hline
\textbf{PSA, median (IQR)} & 399 (98 - 914) & 139 (41 - 412) & 120 (37 - 354) \\
\textbf{LDH, median (IQR)} & 293 (225 - 397) & 235 (188 - 321) & 222 (187 - 308) \\
\textbf{ALP, median (IQR)} & 246 (109 - 447) & 126 (83 - 268) & 125 (79 - 254) \\
\textbf{Hb, median (IQR)} & 11.1 (10.1 - 12.5) & 12.0 (10.8 - 12.8) & 11.9 (0.9 - 12.9) \\
\textbf{Testosterone, median (IQR)} & 11.5 (5.6 - 20.0) & 12.0 (5.7 - 20.0) & 13.0 (5.8 - 20.2) \\ \hline
\end{tabular}%
}
\end{table}
\renewcommand{\arraystretch}{1}

The objective of this analysis was to compare the performance of alternative methods described above in a real-world context in which data were available from a clinical trial and a pseudo EHR dataset, which was constructed from the standard-of-care arm from the clinical trial. We compared the effect of abiraterone acetate (an androgen synthesis inhibitor) plus prednisone compared to prednisone alone on overall survival in patients with metastatic castration-resistant prostate cancer progressing after chemotherapy. Metastatic castration-resistant prostate cancer (MCRPC) is a type of advanced prostate cancer that no longer completely responds to treatments that lower testosterone. \cite{tucci_metastatic_2015} The study sample included patients with MCRPC from a phase 3 randomized double-blind clinical trial (NCT00638690) conducted by Janssen Research $\&$ Development, L.L.C.. Complete details of trial eligibility and treatment protocols have been previously published. \cite{de_bono_abiraterone_2011} 

The clinical trial population included 1,185 patients with MCRPC progressing after taxane chemotherapy. \cite{noauthor_yoda_nodate} Patients were randomized in a 2:1 ratio to treatment with abiraterone acetate plus prednisone or treatment with prednisone alone. Patients were enrolled from 2008 to 2009 and followed for five years, or until death. Treatment arm was classified based on the arm to which a patient was randomized, regardless of whether they crossed over to open-label abiraterone acetate at any point (Table \ref{tab:mcrpc_tab1}).

\begin{figure}
    \centering
    \includegraphics[width=6in]{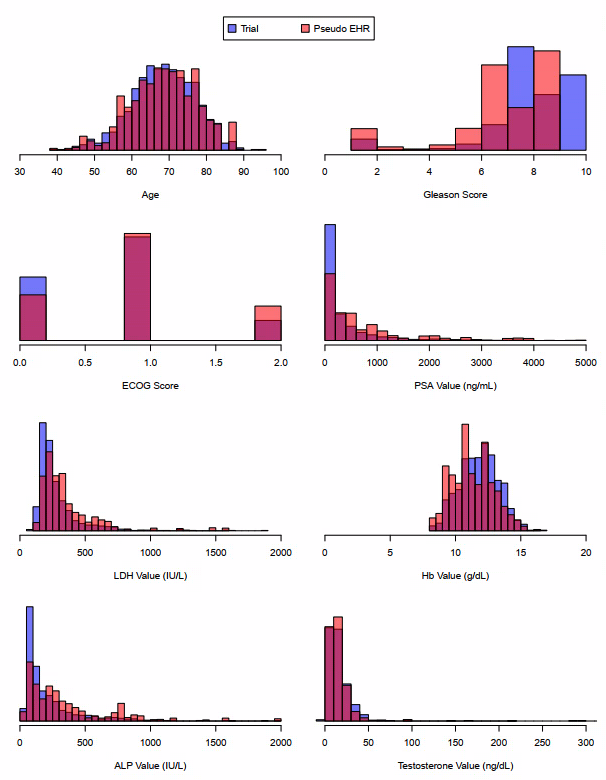}
    \caption{Covariate distributions in the clinical trial and pseudo EHR.}
    \label{fig:real_data_cov_dists}
\end{figure}

The pseudo EHR dataset was constructed by sampling patients from the standard-of-care arm in the clinical trial such that the baseline covariate distribution of the resultant sample differed between the EHR and clinical trial (Table \ref{tab:mcrpc_tab1}, Supplemental Figure \ref{fig:real_data_cov_dists}). We assume that all patients had the same set of covariates associated with poor performance for patients with MCRPC recorded at the baseline encounter. \cite{sorensen_performance_1993, egevad_prognostic_2002, kan_prognosis_2017, mori_prognostic_2019, sharma_alkaline_2014, nakasian_effects_2017} Specifically, all standard-of-care arm patients were sampled with replacement to create a population of size 10,000 from which we could draw our pseudo EHR sample. Next, each patient was assigned a probability of sampling according to a non-linear function of the baseline covariates. By constructing this sampling probability using a non-linear functional form, the estimated on-trial score in the DAW approach will be mis-specified. This reflects the real-world scenario where we are unlikely to be able to correctly specify this model. A psuedo EHR sample of size 1,000 was then drawn. The sampling probabilities were generated such that patients were more likely to be included in the psuedo EHR if they: were younger, had a higher ECOG score, had a higher Gleason score, had a higher lab value for PSA, LDH, Hb, and ALP, or had a lower testosterone value. The logistic sigmoid function was used to relate the ECOG and Gleason scores to the sampling probability in order to separate out those with high versus low scores rather than having alinear additive effect as the score increased. The square root of the testosterone lab value was used to shrink the effect of a high testosterone value on being included in the sample. 

Due to missingness in some variables, multiple imputation via predictive mean matching was used with 5 imputations. The median of the imputed covariates was calculated across imputations and included in the on-trial score, which is valid in the case where the covariates do not inform treatment assignment such as in a clinical trial. \cite{mitra_comparison_2016} Post-imputation covariate distributions stratified by data source were similar to pre-imputation covariate distributions.

\subsection{Case Study Results}

In order to appropriately interpret the results of the case study we first must evaluate each of Pocock's six criteria for the external standard-of-care data. \cite{pocock_combination_1976} In this case meeting most of the criteria was trivial due to the fact that the pseudo EHR was created from the standard-of-care arm of the clinical trial, except for $\#6$, since the pseudo EHR had patients selected in a biased fashion such that their covariate distribution was slightly different and their outcomes were somewhat worse than the clinical trial.

The case study of MCRPC patients had a high rate of death. Of the 791 intervention arm patients in the trial there were 645 deaths, of the 394 patients receiving the standard-of-care in the trial there were 331 deaths, and of the 1000 patients in the pseudo EHR there were 937 deaths. The median survival time was 15.6 months (95$\%$ CI: 14.7-16.8) for the intervention arm in the trial , 11.2 months (95$\%$ CI: 10.4-13.3) for the standard-of-care arm of the trial, and 8.0 months (95$\%$ CI: 7.9-8.6) for the pseudo EHR. It is clear that the patients in the pseudo EHR had inferior survival relative to both arms of the clinical trial. This is likely to be true in reality as patients often receive more supportive care in a clinical trial than in regular clinical practice and tend to have different covariate distributions due to restrictive inclusion/exclusion criteria.

\begin{figure}[ht!]
    \centering
    \includegraphics[width=6.5in]{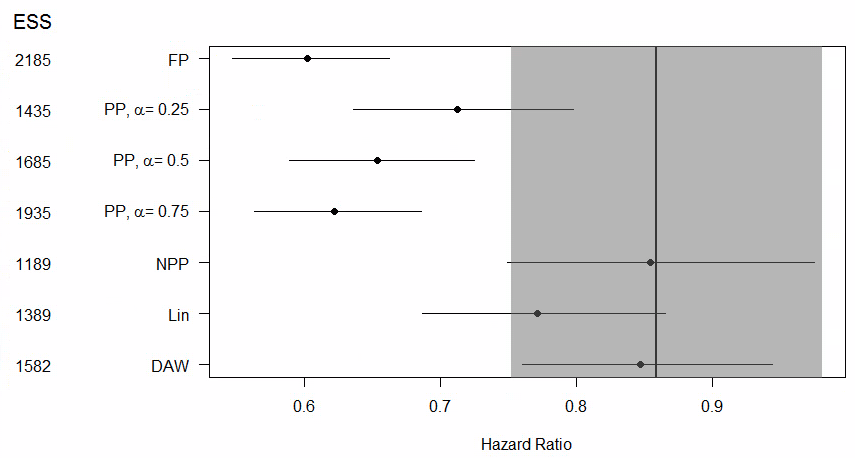}
    \caption{Hazard ratio for death and 95$\%$ confidence intervals (CI) for patients on abiraterone acetate plus prednisone compared to patients on prednisone alone. FP = full pooling, PP = power prior, NPP = normalized power prior ($\hat{\alpha} = 0.004$), DAW = data-adaptive weighting. Vertical line denotes hazard ratio estimated with only the trial data and the grey band denotes the 95$\%$ confidence interval when only the trial data is used. ESS = estimated sample size.}
    \label{fig:yoda_results}
\end{figure}

The hazard ratio for death for patients on abiraterone acetate plus prednisone as compared to patients on prednisone alone was 0.86 (95$\%$ CI: 0.75-0.98) using only patients enrolled in the clinical trial. When all pseudo EHR patients were added to the analysis population, the hazard ratio for death was 0.60 (95$\%$ CI: 0.55-0.66) (Figure \ref{fig:yoda_results}). The hazard ratio for death as estimated by the power prior method with the three different $\alpha$ values were between the trial-only and full pooling methods, as expected. The normalized power prior method estimated $\alpha = 0.004$ and therefore was virtually identical to the trial only method as it effectively borrowed information from only four registry patients. Lin's method returned results somewhat similar to the trial only method, with an estimated hazard ratio of 0.76 (95$\%$ CI: 0.68-0.86), adding just over 200 patients to the analysis (Figure \ref{fig:yoda_results}). The estimated hazard ratio using DAW was 0.85 (95$\%$ CI: 0.76-0.94), which is almost identical to that obtained with the trial only data and also had a smaller confidence interval due to the fact that 397 patients were added so that the augmented trial had a 1:1 randomization ratio (Figure \ref{fig:yoda_results}). With the exception of NPP, which only added 4 patients, DAW returned results most similar to the trial-only result while achieving improved efficiency (Figure \ref{fig:yoda_results}).

\begin{figure}[ht!]
    \centering
    \includegraphics[width=5in]{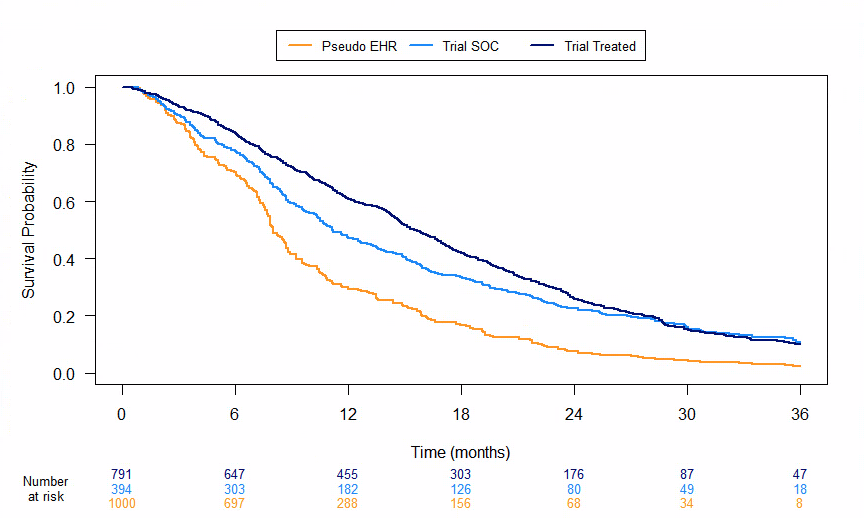}
    \caption{Kaplan-Meier curves for overall survival patients in the clinical trial and the pseudo EHR.}
    \label{fig:yoda_km}
\end{figure}

\section{Discussion}

Data-adaptive weighting allows for the external patients who are most similar to the clinical trial to be selected and weighted such that the augmented trial has a 1:1 randomization ratio, which results in minimal bias and tighter confidence intervals as compared to using only the trial data. We compared the performance of alternative methods for constructing and analyzing a hybrid control arm in terms of bias, variance, power, confidence interval coverage, and type I error across various scenarios for trial size, trial randomization ratio, strength of covariate effects, and treatment effect. 

Based on the results from our simulation study it is clear that fully pooling external patients with trial patients has the potential to produce highly biased results, have poor coverage, and substantially inflate type I error rates. Similarly, the power prior at all three alpha values examined exhibited poor performance, although results were attenuated towards the trial only analyses. The normalized power prior either exhibited moderate bias and poor coverage and type I error rate or had little bias but failed to incorporate much information from the EHR database. The case study examined here shows how the methods perform on real data when Pocock's criteria are met when creating a hybrid control trial. While observed covariates can be accounted for using on-trial scores as in Lin's method and DAW, the other criteria are extremely important to avoid confounding by unobserved characteristics of patients or their care environment, and creation of a hybrid control arm is not recommended if they are not met.

Lin's method has reduced bias and variance compared to the more traditional methods, though DAW was able to achieve lower bias and variance than Lin's method. Both methods performed very similarly with regards to confidence interval coverage and type I error rates. While this may initially cause one to conclude that Lin's method and DAW are both good methods to use for hybrid control arms, there are several points to be made regarding Lin's method. First, Lin's method becomes more difficult to implement with larger trial and/or external data source sizes as optimal matching can be cumbersome or even impossible with large samples. Second, while the number of EHR patients selected by Lin's method nominally produces a 1:1 ratio, weighting by the on-trial score results in an effective sample size that has fewer standard-of-care than intervention arm patients, reducing efficiency of this approach. Finally, it is unclear what estimand Lin's method estimates as the external patients are weighted by their on-trial score and the trial patients are given a weight of 1. These issues are addressed by the DAW as the IOW are scaled to ensure that the 1:1 ratio is preserved and the use of IOW allows one to estimate the estimand of interest: the average treatment effect for those on-trial.

While our simulation study evaluated a large combination of possible characteristics of a trial and real-world data source, there are additional factors that could have an effect on the results that were not examined, including differential covariate effects on the outcome between EHR and trial patients and potential differential error in outcome ascertainment between EHR and trial patients. Furthermore, due to the computational demands of Bayesian estimation for these methods, we have evaluated performance of a frequentist analogue, which targets a different estimand and may produce different results from a Bayesian implementation, particularly for small sample sizes where the Bayesian central limit theorem has little effect. One must also consider the direction of the bias induced by covariates that differ between the trial and real-world populations in order to determine the effect of the differential covariate distributions on power and type I error. 

Based on the results of these simulations and the real-world data example, when working with hybrid-control arm data we recommend using the DAW method in order to minimize bias and variance while maximizing coverage and properly controlling the type I error rate. Additionally, DAW estimates the average treatment effect for those on-trial, which is the estimand of interest.

\section{Acknowledgements}

This study, carried out under YODA Project $\# 2020-4453$, used data obtained from the Yale University Open Data Access Project, which has an agreement with JANSSEN RESEARCH $\&$ DEVELOPMENT, L.L.C.. The interpretation and reporting of research using this data are solely the responsibility of the authors and does not necessarily represent the official views of the Yale University Open Data Access Project or JANSSEN RESEARCH $\&$ DEVELOPMENT, L.L.C.. 

\section{Data Availability}

The data that support the findings of this study are available from JANSSEN RESEARCH $\&$ DEVELOPMENT, L.L.C. via the Yale University Open Data Access Project. Restrictions apply to the availability of these data, which were used under license for this study. Data are available at \url{https://yoda.yale.edu/} with the permission of the Yale University Open Data Access Project.

\section{Funding}
Research reported in this publication was supported in part by NIH grant R21CA227613. The content is solely the responsibility of the authors and does not necessarily represent the official views of the National Institutes of Health.

\section{Declaration of conflicting interests}

The author(s) declared no other potential conflicts of interest with respect to the research, authorship, and/or publication of this article.

\bibliographystyle{unsrt}
\bibliography{Harton_DAW_Manuscript}

\newpage
\section{Supplement}
\beginsupplement

\begin{figure}[ht!]
    \centering
    \includegraphics[width=5in]{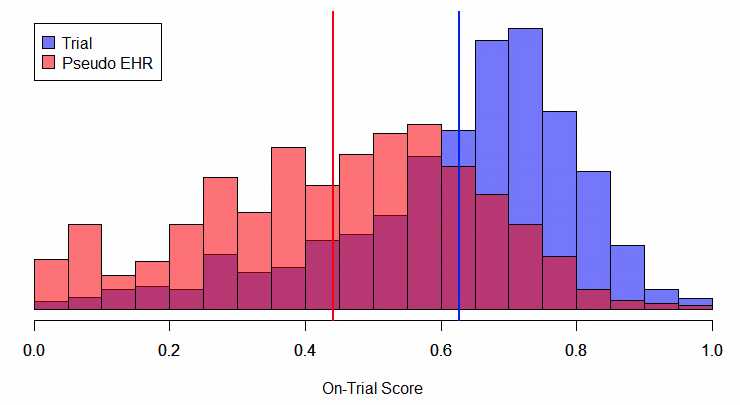}
    \caption{On-trial score distributions for patients in the clinical trial and the pseudo EHR. Vertical lines denote mean on-trial score in each group.}
    \label{fig:ps_dists}
\end{figure}

\end{document}